\documentclass{aa}
\usepackage{graphics}
\begin{document}
\thesaurus{1(11.06.1; 11.09.2; 13.09.1}
\title{A Deep 12 Micron Survey with ISO \thanks{Based on observations with ISO,
an ESA project with instruments funded by ESA Member States (especially the
PI countries: France, Germany, the Netherlands and the United Kingdon) and
with the participation of ISAS and NASA}}    

\author{D.L. Clements$^1$, F-X. Desert$^{1,2}$, A. Franceschini$^3$, W.T. Reach$^{1,4}$,
A.C. Baker$^5$ J.K. Davies$^6$ \and C. Cesarsky$^5$}

\institute{$^1$Institut d'Astrophysique Spatiale, Batiment 121,
Universite Paris XI, F-91405 Orsay CEDEX, France\\
$^2$Laboratoire d'Astrophysique, Observatoire de Grenoble, BP 53, 
414 rue de la piscine, 38041 Grenoble CEDEX 9,  France
\\
$^3$Dipartimento di Astronomia, Universita' di Padova, Padova, Italy\\
$^4$IPAC, Caltech, MS 100-22, Pasadena, CA 91125, USA\\
$^5$Service d'Astrophysique, Orme des Merisiers, Bat. 709, CEA/Saclay,
F91191 Gif-sur-Yvette CEDEX, France\\
$^6$JACH, 660 N. Aokoku Place, University Park, Hawaii, HI 96720, USA\\
}

\offprints{D.L. Clements, Institut d'Astrophysique Spatiale, Batiment 121,
Universite Paris XI, F-91405 ORSAY Cedex, France}
\authorrunning{D.L. Clements et al.}
\titlerunning{A Deep 12 Mircon Survey with ISO}
\date{Received; accepted}

\maketitle
\markboth{D.L. Clements et al.}{Deep 12 Micron Survey with ISO}

\begin{abstract}
We present the first results of a study of faint 12$\mu$m sources
detected with the Infrared Space Observatory (ISO) in four deep, high
galactic latitude fields.  The sample includes 50 such sources in an
area of 0.1 square degrees down to a 5$\sigma$ flux limit of $\sim
500\mu$Jy. From optical identifications based on the US Naval
Observatory (USNO) catalogue and analysis of the optical/IR colours
and Digital Sky Survey (DSS) images, we conclude that 37 of these
objects are galaxies and 13 are stars. We derive galaxy number counts
and compare them with existing IRAS counts at 12$\mu$m, and with
models of the 12$\mu$m source population. In particular, we find
evidence for significant evolution in the galaxy population, with the
no-evolution case excluded at the 3.5$\sigma$ level. The
stellar population is well matched by existing models.  Two of the
objects detected at 12$\mu$m are associated with known galaxies. One
of these is an IRAS galaxy at z=0.11 with a luminosity of 10$^{11}
L_{\odot}$.
\keywords{galaxies:evolution, galaxies:infrared, stars:infrared}
\end{abstract}

\section{Introduction}

Whenever a radically new wavelength or sensitivity regime is opened in
astronomy, new classes of object and unexplained phenomena are
discovered. The IRAS satellite, in opening the mid- to far-IR sky,
revealed  that the bolometric luminosities of many galaxies are
dominated by this spectral region. This has raised questions
concerning the evolution of galaxies in the mid- to far-IR, the role
of dust-extinction in the formation of galaxies, the
relationship between quasars and galaxies, and the nature of galaxy
formation itself. 
Limited as they were to fluxes not much smaller than 1Jy, the 
 IRAS surveys were constrained to the fairly local
universe for the vast majority of the detected objects. 
The evolutionary properties of IR galaxies, 
both the normal galaxy population and the unusual
 ultra- and hyper-luminous objects (see e.g. Clements et al.
1996a), are thus still mostly unknown.

Recent work in the visible and near-IR has had a major impact on our
understanding of the star formation history of galaxies (eg. Steidel
et al., 1996, Cowie et al., 1994, Giavalisco et al., 1996, Lilly et
al., 1996). It appears that the star formation rate in the universe
peaked at around z=1 and has been declining since (Madau et al. 1998).
Many of the objects studied in deep, high redshift fields appear to
be distorted, and are possibly undergoing tidal interaction or merging
(Abraham et al., 1996). It is well--known that tidal interactions and mergers
in the local universe produce significant amounts of star formation
(Joseph \& Wright, 1985, Clements et al., 1996b, Lawrence et al., 1989), and
that these are usually associated with a significant luminosity in the
mid- to far-IR. The dust responsible for this emission is heated by
the star formation process, which it also obscures. We must thus
consider that the view of the universe obtained in the visible and
near-IR, corresponding to the rest-frame optical and near-UV of many
of the objects observed, may well be biased by such obscuring
material. The question of how much of the star-formation in the
universe is obscured by dust thus becomes important.  This issue can
only be properly addressed by selecting objects in the mid- or far-IR
which are less affected by such obscuration.

Previous work in the mid- and far-IR used data from IRAS,
with all-sky sensitivities of $\sim$0.1 Jy in the mid-IR bands (12 and
25 $\mu$m), and $\sim$ 0.3--1 Jy at 60 and 100 $\mu$m. These typically
allow the detection of galaxies out to z=0.2, though a few exceptional
objects, such as the gravitationally lensed z=2.286 galaxy
IRAS10214+4724 (Rowan-Robinson et al., 1991, Serjeant et al., 1995),
have also been detected.

Most evolutionary studies with IRAS have concentrated on the 60$\mu$m
waveband (Sanders et al., 1990, Hacking \& Houck, 1987, Bertin et al.,
1997). This work has found evidence for strong evolution in the 60$\mu$m
population, at rates similar to those of optically selected AGN, but
the nature of this evolution is still unclear, and it is difficult to
extrapolate to higher redshift with any confidence.

At mid-IR wavelengths, the IRAS mission has produced both large-area
surveys of fairly nearby objects (eg. Rush et al., 1993), and
small-area, deep surveys in repeatedly scanned regions (eg.  Hacking
\& Houck, 1987). The former surveys do not probe sufficiently deeply
into the universe to be able to say much about galaxy evolution, but
they do have the advantage that plentiful data exists for the nearby
galaxy samples they produce. The latter surveys are plagued by stellar
contamination. The vast majority of the 12$\mu$m objects in the survey
of Hacking et al., for example, are stars -- there are only five
galaxies in their entire survey.

The Infrared Space Observatory (ISO, Kessler et al. 1996) provides a
major improvement to our observational capabilities beyond those of
the IRAS satellite. For observations in the mid-IR, the ISOCAM
instrument (Cesarsky et al. 1996) allows us to reach flux limits $\sim$100
times fainter than those achieved by IRAS while observing fairly large
areas ($\sim$0.1 sq. degree) in integration times of only a few
hours. We can thus probe flux regimes that were previously impossible
to study.

This paper presents the results of a survey of four high
galactic latitude fields using the LW10 (12$\mu$m) filter, which was
specifically designed to match the 12$\mu$m filter on the IRAS
satellite.  The present results can thus be compared to existing IRAS
data with minimal model-dependent K- corrections.
This survey is much deeper than any based on IRAS data, and is
sufficiently deep to detect distant galaxies. The survey region is
also fairly small and at high galactic latitude, so that stellar
contamination should not be a major problem.

There are of course other studies in the mid-IR underway using the ISO
satellite. These include the DEEP and ULTRADEEP surveys (Elbaz et al.,
in preparation), the ELAIS survey (Oliver et al., in preparation) and
the ISOHDF project (Oliver et al., 1997; Desert et al., 1998 (Paper
I); Aussel et al. 1998).  All of these programmes use the LW2
6.7$\mu$m and/or LW3 15$\mu$m filter on ISO. Only the ISOHDF results
have been published to date. At 15$\mu$m these observations reach a
flux limit of $\sim$0.1 mJy, about 5 times deeper than the
observations discussed here, but cover only 1/24th of the area.
There are also deep surveys at longer wavelengths using the PHOT
instrument at 175$\mu$m (Kawara et al., 1998, and Puget et al., 1998).
Future missions will also be probing this part of the electromagnetic
spectrum. The first of these will be the WIRE mission (Fang et al.,
1996) which will obtain a large area, deep survey at 12 and
25$\mu$m. The SIRTF project (Werner \& Bicay, 1997) and IRIS satellite
(Okuda, 1998) will also be used for deep number counts, and should be
able to probe significantly deeper than ISO. Finally the planned NGST
(Stockman \& Mather, 1997) will provide incomparable performance in
this cosmologically interesting waveband.  The present work provides
the first results of the exploration of the distant universe at these
mid-IR wavelengths, and can provide a guideline for future missions,
useful for their planning and preparation.

The paper is organised as follows. Section 2 describes
the observations, data reduction and calibration. Section 3 provides
details on identifications of the 12$\mu$m source population, 
star-galaxy separation, and on individual source properties. 
Section 4 discusses number counts, comparison with statistics 
at other wavelengths, and with model predictions.
Section 5 summarises our conclusions.

\section{Observations and Data Reduction}

The observations presented here were part of a survey of cometary dust
trails (ISO project name JDAVIES/JKDTRAIL).  The original goal was to
observe the width and structures of the cometary trails, which are
produced by large particles, ejected from the comet into independent
heliocentric orbits but with very similar orbital elements and very
small radiation pressure effects. The comet 7P/Pons-Winnecke was
selected because of its similarity to other comets with dust trails,
the detection of a dust trail by IRAS (Sykes \& Walker, 1992) and its
suitability for observation by ISO. Four fields were imaged, each
field being a raster map centred on the ephemeris prediction for
particles with the same orbital elements as the nucleus except for the
mean anomaly of the orbit, which was shifted by $+1^\circ$ (ahead) and
$-0.5^\circ$, $-1^\circ$, and $-2^\circ$ (behind).  Each raster was 11
by 7 pointings, with a spacing of $60^{\prime\prime}$ by
$48^{\prime\prime}$. The pixel field of view was $6^{\prime\prime}$,
so that there was substantial overlap between individual frames of the
32 by 32 pixel ISOCAM LW array (Cesarsky et al., 1996) to ensure good
flatfielding and high observational redundancy.  A typical position on
the sky was visited 12 times during the raster, each time with a
different pixel of the array.  The rasters were rotated such that the
predicted cometary trail would run parallel to the short axis of the
raster. Unfortunately for the cometry trails programme, the
observations took place on 17 August 1997, one day later than assumed
for the ephemeris calculations and therefore the trail is predicted to
run horizontally across the very bottom edge of the image. Based on
the results of observations of the trail of another comet (P/Kopff)
from the same observing program (Davies et al., 1997), we expect that
the trail would occupy at most the lower $1^{\prime}$ of the image and
that it would be relatively smooth. The presence of the dust trail in
the field should not have any effect on the observations presented
here, though an excess of sources in the bottom of the raster could
potentially be related to structures in the dust trail. No such excess
is seen. All observations presented here are in the LW10 filter, which
is very similar to the IRAS 12 $\mu$m filter in
wavelength-dependent response. Since these fields are at high galactic
lattitude ($>$50 degrees), and, in the absence of cometary trails, are
effectively blank fields, they become ideal for a deep survey of the
extragalctic 12$\mu$m source population. The positions of the fields are
given in Table ~\ref{fields}.

\begin{figure*}
\resizebox{\hsize}{!}{\includegraphics{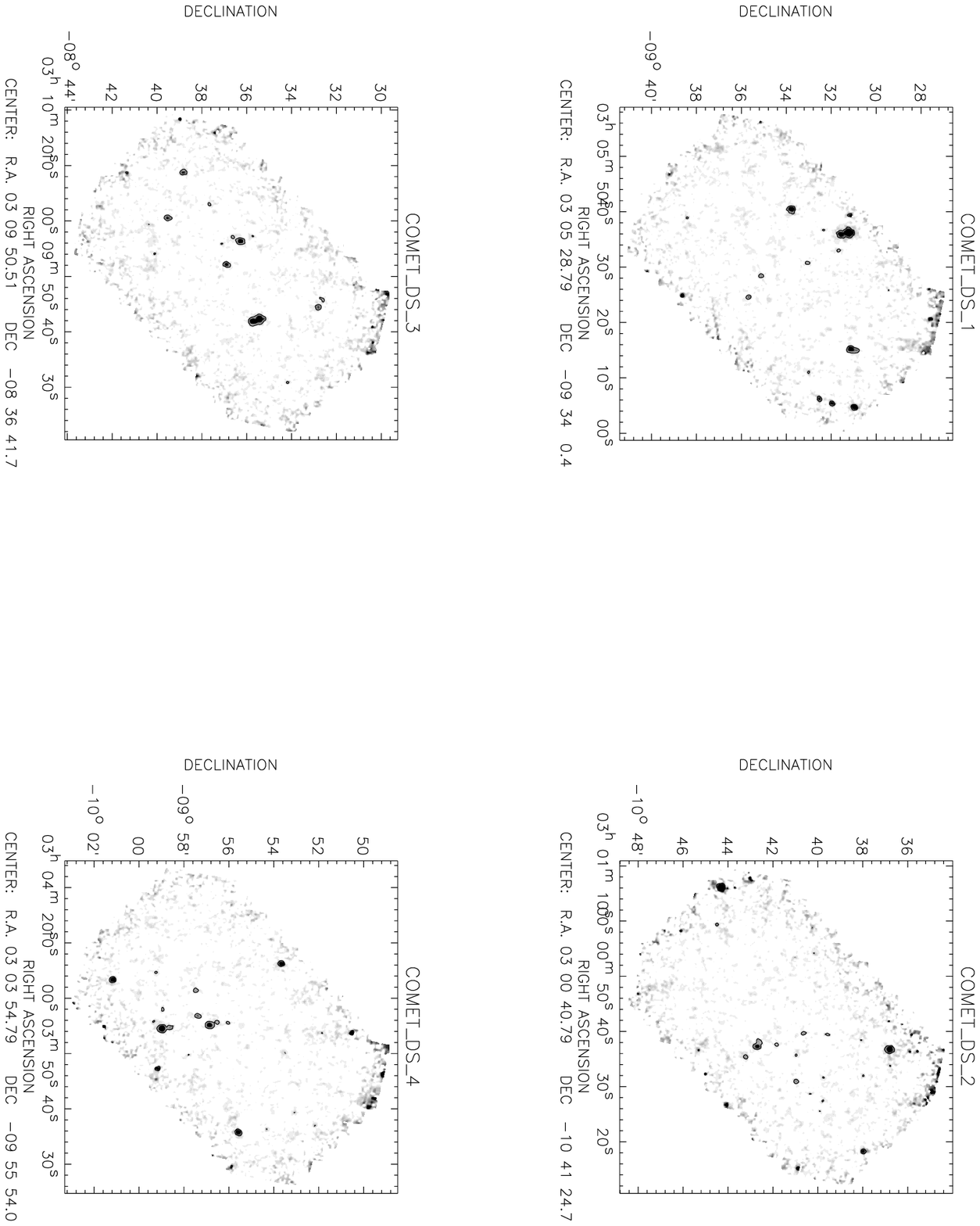}}
\caption{Maps of the final reduced ISOCAM data
for the four 12$\mu$m survey fields}
\label{images}
\end{figure*}

The data reduction process is described in detail in Paper
1. Basically, for each AOT file (total 4) the raw data (CISP format)
and pointing history (IIPH format) are read and merged.  The raw data
cube (typically 1244 readouts of 32 by 32 pixels) is deglitched for
fast and slow cosmic ray impacts. A transient correction is applied to
recover the linearity and the nominal flux response of the camera. A
triple--beam method is then applied to the processed data cube, in
order to find the best (ON- (OFF1+OFF2)/2) differential value of the
sky brightness for each pixel and each raster position, where OFF1 and
OFF2 refer to the previous and next raster position value for the same
camera pixel. The resulting low--level reduced cube is then simply
flat--fielded (there is no need for a dark correction because we
perform a differential measurement on the sky). The flat--fielded
reduced cube along with an error cube is made up of 77 values for each
of the 32 by 32 pixels of the camera. The 2 cubes are then projected
onto the sky using an average effective position for ISO during each
raster position, with a $1/\sigma^2$ optimal weighting. A noise map is
thus calculated as well as a sky differential map. Any given source
will leave 2 negative half--flux ghosts 60 arcseconds away on both
sides along the raster scan direction. This is a trade--off in order
to beat the 1/f noise regime that is reached by the camera for long
integrations per position.  On the final map, shown in
Fig. ~\ref{images}, we search for point sources with a Gaussian
fitting algorithm that uses the noise map for weighting the pixels and
deducing the noise of the final flux measurement. A FWHM of 9
arcseconds was used. The final internal flux is converted to
$\mu$Jy by using the ISOCAM cookbook value (Cesarsky et al., 1994).
The present understanding of ISOCAM calibration is that no additional
factor should be applied to the pre-flight sensitivity estimates for
the determination of surface brightness.

We have devised a scheme to assess the reproducibility of the sources,
in order to test for false sources that would be due to undetected
glitches. This is described in detail in Paper I, but consists of
breaking the observations of each object into a number of independent
subsurveys. Sources are deemed reliable if they are detectable, with
suitably reduced significance, in each of these subsurveys. Of the 193
sources above the 3$\sigma$ limit only 7 fail the reproducibility test
(4\%) and these are not considered in the following. Visual screening
helped in removing a further 38 residual companions of strong sources
due to imperfect fitting. Visual screening also showed that two
sources, F1\_0 and F2\_0, were significantly extended at 12$\mu$m. We
thus use aperture photometry to obtain an accurate flux for these
objects.  The aperture used had a diameter of 20 arcsecs.

Simulation of the expected PSF from ISO after the same processing
reveals that part of the flux is missed by the optimised Gaussian
fitting. We therefore correct the fluxes and errors by a factor of
1.52 determined from this modelling. The final absolute photometry
should be in error by no more than an estimated 30\%.  The fluxes are
given at the nominal wavelength of 12 $\mu$m (i.e. an additional
correction of $1.04= 12/11.5$ is applied since the nominal ISOCAM
calibration is for a wavelength of 11.5$\mu$m), in order to facilitate
the comparison with previous IRAS observations. This assumes a flat
spectrum in $\nu F_{\nu}$ as was used for IRAS calibration.  The flux
prediction for known stars should thus be colour corrected, since they
have a Rayleigh-Jeans spectral index, in order for comparison to
ISOCAM measurements: the real flux at 12$\mu$m should be divided by
0.902. An additional factor comes from the fact that the PSF is
smaller for stellar sources (which are dominated by the short
wavelength part of the broad filter) than for the assumed
extragalactic sources which have a broader spectrum.  Thus the real
flux should also be multiplied by a supplementary factor of 1.13 (see
Section 3.1 for this {\it a priori} calibration and the comparison with
flux measurements of known stars in the fields). The basic calibration of
ISOCAM, before the corrections for point sources are applied, can be
checked by comparing the integrated surface brightness of these fields
with values interpolated from the DIRBE experiment on COBE (Hauser et
al. 1997a). The ISO surface brightnesses agree with the DIRBE values
to better than 5\%.

The sensitivity that is achieved in the central area of the fields is
about $1\sigma = 100 \mu$Jy. This was achieved with a total
integration time of 4 minutes for each camera field--of--view.
Astrometry was assessed by matching ISO sources to bright stars in
the fields. We estimate the astrometric accuracy to be $\sim$6'' (2 $\sigma$).

\begin{table}
\begin{tabular}{lrrrr}
Field&RA&Dec&l&b\\ \hline
Field 1&03 05 30&-09 35 00&190.327&-53.895\\
Field 2&03 00 40&-10 42 00&190.813&-55.466\\
Field 3&03 09 50&-08 37 00&189.979&-52.433\\
Field 4&03 04 00&-09 56 00&190.477&-54.361\\
\end{tabular}
\caption{Details of 12$\mu$m Survey Fields}
RA and Dec are in J2000. All fields have high galactic latitude $>$50
degrees, making them ideal for cosmological studies.
\label{fields}
\end{table}

\section{The 12$\mu$m Source Population}

A total of 186 candidate 12$\mu$m sources are found in the survey to a
3$\sigma$ flux limit of $\sim$ 300$\mu$Jy. Visual inspection then
removes 38 of these sources as fragments of brighter sources
incorrectly identified as separate objects, giving a master list of
148 objects. For the remainder of this paper we shall restrict
ourselves to discussion of only those sources detected at 5$\sigma$
sensitivity or above in this master list. This is for several reasons.
Firstly, a number of uncertainties remain in the identification of the
weakest sources. These are the sources most likely to be affected by
the remnants of subtracted glitches or by weak, undetected
glitches. Further examination of the detailed time histories and
reproducibilities of these sources is underway, and a full catalogue
reaching to the faintest flux limits can then be
constructed. Secondly, the problems of Malmquist bias (Oliver, 1995)
are most easily controlled in catalogues detected with significances
$\geq 5\sigma$.  A source list using a 5$\sigma$ detection threshold
is thus best suited to our examination of the 12$\mu$m source
counts. 50 objects are detected at 5$\sigma$ or greater
significance. Details of these objects are given in Table
\ref{catalog}, and they are discussed further in the following sections.

\begin{table*}
\begin{tabular}{rrrrr}
Name&F$_{12}$($\mu$Jy)&RA(2000)&DEC(2000)&Class\\ \hline F1\_0&
12147.$\pm$ 200.& 3 5 36.2& -9 31 21.6&Extended; IRAS Source\\ F1\_1&
4258.$\pm$ 219.& 3 5 4.5& -9 31 6.6& Star\\ F1\_2& 4155.$\pm$ 131.& 3
5 40.4& -9 33 56.1& Star\\ F1\_3& 3835.$\pm$ 99.& 3 5 15.1& -9 31
17.6&\\ F1\_4& 2150.$\pm$ 185.& 3 5 5.3& -9 32 7.7&\\ F1\_5&
1046.$\pm$ 102.& 3 5 24.5& -9 35 50.7&\\ F1\_9& 3732.$\pm$ 142.& 3 5
35.9& -9 31 43.8&\\ F1\_10& 1903.$\pm$ 185.& 3 5 6.1& -9 32 41.5&\\
F1\_11& 1324.$\pm$ 113.& 3 5 14.9& -9 31 3.8&\\ F1\_12& 842.$\pm$ 92.&
3 5 28.4& -9 35 17.5&\\ F1\_18& 2190.$\pm$ 171.& 3 5 36.3& -9 31
32.2&\\ F1\_22& 615.$\pm$ 112.& 3 5 33.0& -9 31 50.4& Star\\ F1\_30&
636.$\pm$ 88.& 3 5 30.7& -9 33 13.4&\\ F1\_44& 622.$\pm$ 109.& 3 5
36.7& -9 32 30.1&\\ F1\_48& 605.$\pm$ 119.& 3 5 11.0& -9 33 10.0&\\
F2\_0& 10479.$\pm$ 260.& 3 1 6.1& -10 44 27.5&GSC-galaxy; Extended\\
F2\_1& 6884.$\pm$ 175.& 3 0 36.7& -10 36 57.3& Star\\ F2\_3&
2385.$\pm$ 100.& 3 0 37.3& -10 42 51.0&\\ F2\_6& 911.$\pm$ 99.& 3 0
31.0& -10 41 7.2&\\ F2\_12& 744.$\pm$ 136.& 3 0 59.4& -10 44 38.2&\\
F2\_16& 770.$\pm$ 104.& 3 0 38.2& -10 42 46.6& Star\\ F2\_23&
702.$\pm$ 95.& 3 0 39.7& -10 40 47.0& Star\\ F2\_24& 814.$\pm$ 118.& 3
0 35.4& -10 43 22.3&\\ F2\_80& 522.$\pm$ 89.& 3 0 39.5& -10 39 43.4&\\
F2\_87& 536.$\pm$ 88.& 3 0 37.6& -10 41 59.4&\\ F2\_195& 464.$\pm$
87.& 3 0 35.8& -10 41 6.5&\\ F3\_0& 7265.$\pm$ 104.& 3 9 42.6& -8 35
33.4&Binary Star\\ F3\_1& 3266.$\pm$ 98.& 3 9 56.7& -8 36 24.5& Star\\
F3\_2& 1901.$\pm$ 100.& 3 9 52.5& -8 37 1.5& Star\\ F3\_3& 1760.$\pm$
95.& 3 10 0.9& -8 39 39.2&\\ F3\_4& 1682.$\pm$ 105.& 3 10 9.1& -8 38
57.2&\\ F3\_5& 1374.$\pm$ 108.& 3 9 44.7& -8 32 55.6&\\ F3\_9&
5444.$\pm$ 93.& 3 9 42.2& -8 35 50.3&Binary Star\\ F3\_15& 597.$\pm$
100.& 3 10 3.3& -8 37 46.7&\\ F3\_34& 639.$\pm$ 105.& 3 9 46.1& -8 32
44.0&\\ F3\_37& 648.$\pm$ 101.& 3 9 57.4& -8 36 44.2&\\ F4\_0&
7684.$\pm$ 140.& 3 3 54.5& -9 59 6.7& Star\\ F4\_1& 3670.$\pm$ 149.& 3
4 3.4& -10 1 18.0&\\ F4\_2& 3581.$\pm$ 155.& 3 3 35.8& -9 55 43.1&\\
F4\_3& 3207.$\pm$ 84.& 3 3 55.2& -9 56 59.9&\\ F4\_4& 3166.$\pm$ 357.&
3 3 53.8& -9 50 40.0& Star\\ F4\_5& 2481.$\pm$ 215.& 3 3 47.3& -9 59
18.2&\\ F4\_6& 3051.$\pm$ 197.& 3 4 6.3& -9 53 47.6&\\ F4\_9&
1234.$\pm$ 95.& 3 3 56.8& -9 57 31.3&\\ F4\_11& 806.$\pm$ 89.& 3 4
1.4& -9 57 36.6&\\ F4\_12& 933.$\pm$ 106.& 3 3 54.7& -9 58 45.6&\\
F4\_24& 559.$\pm$ 92.& 3 3 58.0& -9 59 5.2&\\ F4\_28& 474.$\pm$ 88.& 3
3 55.6& -9 56 11.0&\\ F4\_57& 615.$\pm$ 118.& 3 3 55.8& -9 59 28.3&\\
F4\_68& 690.$\pm$ 87.& 3 3 55.7& -9 56 40.4& Star\\
\end{tabular}
\caption{12$\mu$m Sources Detected at $> 5\sigma$}
Objects are classified as stars based on their optical/IR colours (see
section 3.1). The object indicated as GSC-galaxy is in the GSC but
inspection of the DSS image reveals it to be a galaxy. Two of the stars
F3\_0 and F3\_9 are very close, precluding accurate photometry.
\label{catalog}
\end{table*}

\subsection{Optical Identifications and Star-Galaxy Separation} 

\begin{figure}
\resizebox{\hsize}{!}{\includegraphics{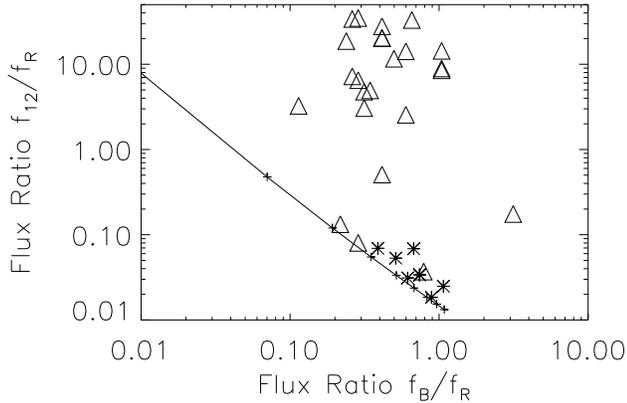}}
\caption{Optical-IR Colour-Colour Diagram}
Triangles are anonymous 12$\mu$m sources, stars are objects identified
with GSC stars, the line shows the locus for Black Bodies, with
temperature decreasing from 10000 K in the lower left. Points on the
line lie at 1000K intervals. Note the concentration of stars on the
Black Body line, and the majority of sources lying well away from the
line. This allows star-galaxy separation.
\label{optplot}
\end{figure}

Comparison of the 12$\mu$m ISOCAM images with Digital Sky Survey
(DSS)\footnote{Based on photographic data of the National Geographic
Society -- Palomar Observatory Sky Survey (NGS-POSS) obtained using
the Oschin Telescope on Palomar Mountain.  The NGS-POSS was funded by
a grant from the National Geographic Society to the California
Institute of Technology.  The plates were processed into the present
compressed digital form with their permission.  The Digitized Sky
Survey was produced at the Space Telescope Science Institute under US
Government grant NAG W-2166.}  images shows that a number of the
sources are associated with bright stars. Before we are able to
analyse the galaxy component of the 12$\mu$m source population, these
and any other contaminating stars must be identified and removed. This
was achieved by using the US Naval Observatory (USNO) all-sky
photometric catalogue (Monet et al., 1996). The database was searched
for all optical objects within 12 arcseconds of each ISO position. 12
sources were immediately identified with HST Guide Star Catalogue
(GSC) stars, though inspection of the DSS images shows that one of
these is in fact a galaxy (03 01 06.16 -10 44 23.6, the GSC `star'
5290\_640). B and R band photometry was extracted for 29 of the 32
optically identified objects -- three of the GSC stars were too bright
to allow photometry from the B survey plates used by the USNO
catalogue. These magnitudes were then corrected for the estimated
galactic extinction. A comparison of the final F$_B$/F$_R$ and
F$_{12}$/F$_R$ flux ratios was then made. Figure~\ref{optplot} shows
the optical/ISO colour-colour diagram, together with a Black Body
colour track. Simple stars, without associated dust or stellar
companions, should lie on or near to this colour track. As can be
seen, almost all of the GSC stars and several other objects lie near
to the Black Body line. This allows us to remove all those stars that
have not been identified in the GSC. Three such objects are
removed. One star (F3\_9: 03 09 42.14 -08 35 44.6) seems to be
anomalously blue (B-R = -0.8 from the USNO catalogue). However, this
object and another bright star (F3\_0: 03 09 42.6 -08 35 33.4) are so
close to one another that accurate photographic photometry is likely
to be difficult, resulting in the anomalous colours. These objects are
removed from further analysis.

Of the 32 optically identified 12$\mu$m sources we thus conclude that
13 are stars and that the remaining 19 are optically identified
galaxies. 18 sources, all probably galaxies, thus remain without
optical identifications to the limits of the USNO-A catalogue
ie. around 20th magnitude in B and R.

The colour-colour plot also allows us to check the calibration for the
12$\mu$m survey.  We can use the B-R colours to provide a rough
spectral type for all stars in the survey. This can then be
cross-referenced to the surface temperature of that stellar type. The
12$\mu$m flux can then be extrapolated from the R band flux, assuming
a simple Black Body spectrum. This approach suggests that the flux
calibration is accurate at the $\sim$20\% level (see
Table~\ref{stars}). We have also checked these results using detailed
spectral energy distributions (SEDs) based on the Kurucz stellar atmosphere
codes instead of a simple Black Body extrapolation, and arrive at very
similar conclusions.  The main source of uncertainly here is the
treatment of undersampled unresolved sources in the ISOCAM reduction
systems. As more data becomes available on the details of the ISOCAM
PSF, this systematic uncertainty will be reduced. There is also the
possibility that one or more of the stars in the survey have genuine
IR excesses. Ground-based near- to mid-IR photometry will be required
to confirm this.

\begin{table*}
\begin{tabular}{rrrrrrr}
Name&F$_{12}$($\mu$Jy)&Bmag&Rmag&F$_{12 PRED}$ ($\mu$Jy)&Pred/Obs&Excess?\\ \hline
F1\_1&4258&12.6&11.7&1794&0.42&Yes\\
F1\_2&4155&10.9&10.3&4962&1.19\\
F1\_22&615&13.1&12.7&442&0.72\\
F2\_1&6884&13.1&11.3&8018&1.16\\
F2\_16&770&13.7&12.7&747&0.97\\
F2\_23&702&13.7&12.9&540&0.77\\
F3\_2&1902&13.5&12.3&1220&0.64\\
F3\_1&3266&13.5&12.0&2289&0.70\\
F4\_68&690&13.7&12.9&720&1.04\\
\end{tabular}
\caption{Predicted and Actual Stellar Fluxes in the 12$\mu$m ISO Survey}
Predicted fluxes are calculated as described in the text. A star is
described as having a candidate IR excess if its predicted flux is less than
half the measured flux.
\label{stars}
\end{table*}

We are then able to remove all 13 stars from the 12$\mu$m source lists
generated in this survey, and can thus examine the statistics of faint 
galaxies at 12$\mu$m.

\subsection{Individual Sources}

We discuss here individual sources of note in this survey.

{\bf IRAS 03031-0943} This lies at 03 05 36.4 -09 31 27.0 (J2000) and
is an IRAS source identified with a B=18.6 galaxy at z=0.112 (Clements
et al., 1996a). It is associated with object F1\_0 in the present
survey. This galaxy has IRAS fluxes of 0.85 and 0.51 Jy at 100 and
60$\mu$m respectively, and limits of 0.15 and 0.095 Jy at 25 and
12$\mu$m, consistent with the measured ISO 12$\mu$m flux of
12.1$\pm$0.2 mJy. This galaxy has a 60$\mu$m luminosity of
10$^{11}L_{\odot}$ (H$_0$=100kms$^{-1}$Mpc$^{-1}$, q$_0$=0.5) which
places it among the high luminosity IRAS galaxies but at lower
luminosity than the ultraluminous class (see eg. Sanders \& Mirabel,
1996). Its optical spectrum contains strong emission from the
H$\alpha$-NII blend and SII, but the redshift measurement spectrum is
of too low a resolution to provide any emission line diagnostics
(Clements, private communication). We thus do not know what sort of
power source is energetically dominant in this object --- starburst or AGN.

{\bf NPM1G-10.0117} This lies at 03 01 06.2 -10 44 24 (J2000) and is a
B=16.63 galaxy used in a proper-motion survey (Klemola et al., 1987). It
is associated with object F2\_0 in this survey, and has a 12$\mu$m
flux of 9.10$\pm$0.26 mJy. It is also identified with HST
Guide Star GSC 5290\_640, but is clearly a galaxy in the Digitised Sky
Survey images. Little else is currently known about it.

Altogether, we can say relatively little about the galaxies identified
so far in this survey since the identification programme has only just
started. Nevertheless, it is a useful check of the processing to note
that the only IRAS galaxy within the survey region has been detected
by ISOCAM.

\section{The 12 $\mu$m Number Counts}

\begin{figure}
\resizebox{\hsize}{!}{\includegraphics{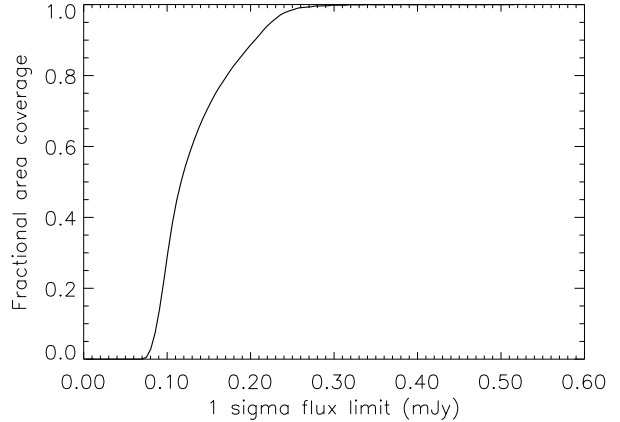}}
\caption{Areal Coverage Plot for 12$\mu$m Survey}
\label{coverage}
\end{figure}

Integral number counts from a survey with homogeneous sensitivity are
calculated by summing up the number of sources
to a given flux limit, and then dividing by the survey  area:
\begin{equation}
\label{eq:rawc}
N(>S) = \sum_{flux>S} \frac{1}{\Omega}
\end{equation}
where $\Omega$ is the area of the survey. However, in our case
the noise in the survey is inhomogeneous (it has a bowl-like shape) since
the border pixels were observed with smaller integration times. We thus have
to make a correction to equation \ref{eq:rawc} to account for this.
If we define $\eta (\sigma )= \Omega/ \Omega(\leq
s )$, where $\Omega(\leq s )$ is the area where the 
$1\sigma$ sensitivity is
better than $s$, then the corrected number counts are given by:
\\
\begin{equation}
\label{eq:cts}
N(>S) = \sum_{flux>S} \frac{1}{\Omega} \times \frac{1}{\eta(S/n)}
\end{equation}
where n is the detection threshold of the survey. In the present paper we 
consider only those sources detected at $> 5 \sigma$ confidence, so n=5.
We plot the area coverage, $\eta$, in Fig. ~\ref{coverage}.

A correction must also be applied to account for Malmquist bias
(Oliver, 1995). This bias arises when looking at number counts for
a population with rapidly increasing numbers at fainter fluxes, as is
the case for our 12$\mu$m galaxy sample. In the presence of
observational noise, some galaxies close to, but below, the flux limit
will be scattered above the flux limit by noise and will appear in the final
catalogue. Similarly some galaxies close to but above the flux limit
will be scattered out of the catalogue.  However, since there are many
more galaxies at fainter fluxes, more galaxies will be scattered above
the flux limit than below it. Number counts that are uncorrected for
this bias thus show a steep rise in counts towards the faintest flux
levels. In the case of Gaussian noise and a Euclidean count slope,
which approximates to the present case, Murdoch et al (1973) tabulated
the effects of this bias to a detection level of 5$\sigma$, allowing for the
observed fluxes to be corrected. Oliver (1995) provides a
numerical version of this correction which we apply here. For
observations probing below the 5$\sigma$ limit this simple correction
cannot be applied, and a more complex Monte Carlo approach must be
adopted (eg. Bertin et al. 1997).

\begin{figure*}
\resizebox{\hsize}{!}{\includegraphics{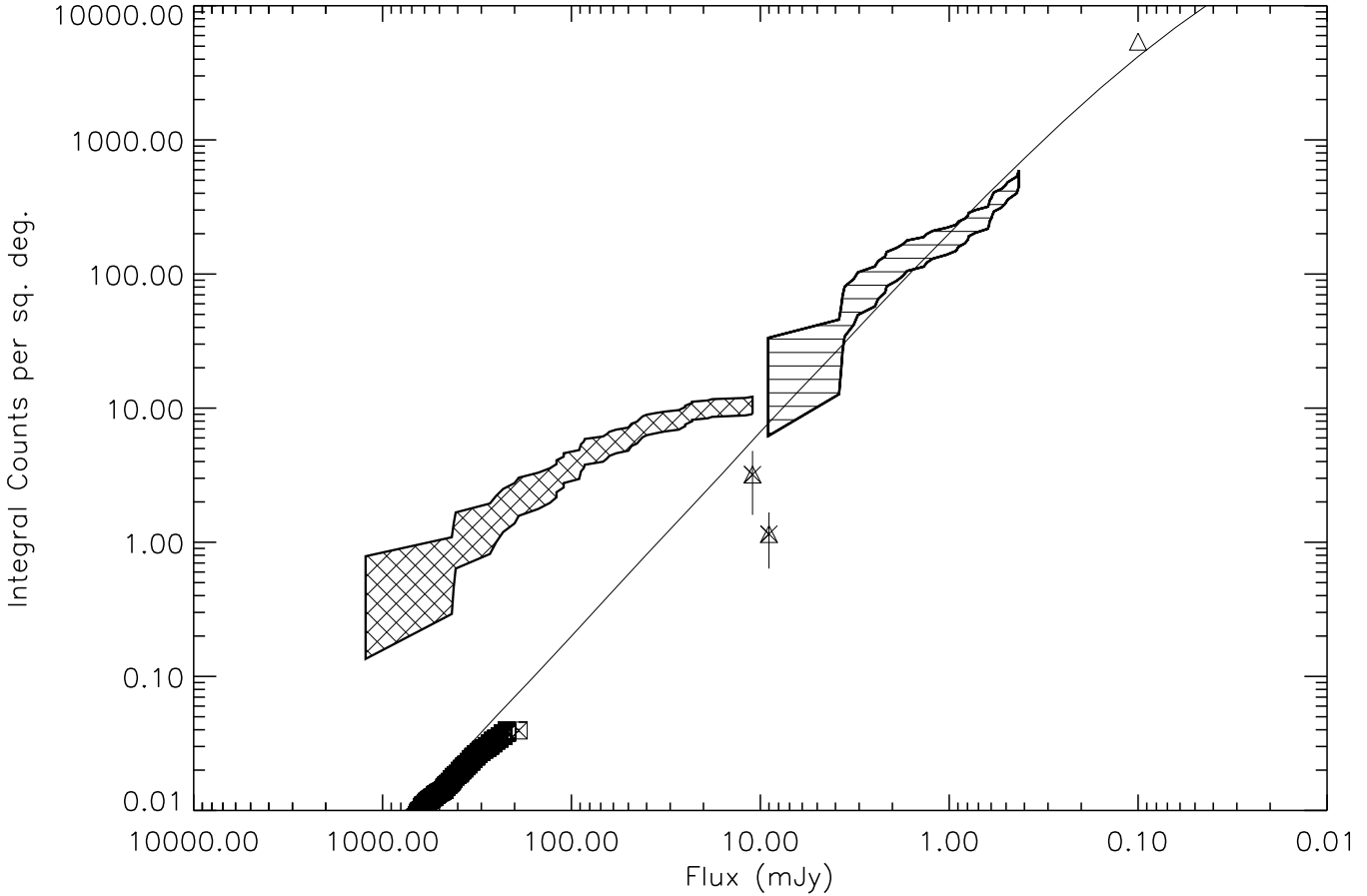}}
\caption{12$\mu$m Integral Galaxy Counts from ISO}
This plot presents the current observational state of knowledge on
galaxy counts at 12$\mu$m. Since the errors in integral-counts plots
are correlated, we show the areas allowed within the $\pm 1\sigma$
error bars as shaded regions.  The cross-hatched region indicates the
deep IRAS counts from Hacking \& Houck (1987) including stars.
Triangles indicate the galaxy content of this survey (lower) and the
Gregorich et al. survey (1995) (upper). The solid region in the bottom
left shows the counts from the extended IRAS galaxy survey by Rush et
al. (1993) using the all-sky IRAS Point source Catalogue, with the
area of this survey taken from Oliver et al. (1997).  The
single-hatched region shows the results of the present work, reaching
fainter magnitudes than previous 12$\mu$m studies. The 13 stars among
the 50 objects detected at $>5\sigma$ are not included in these count
statistics. The triangle in the top right shows the ISOHDF counts at
15$\mu$m (Oliver et al. 1997) -- note that no correction has been
applied to this data to convert from 15 to 12 $\mu$m.  The solid line
is the strong evolution model which is discussed in the text.
\label{intcounts}
\end{figure*}

Figure~\ref{intcounts} shows the Malmquist-bias corrected integral
number count plot from the present work and from 12$\mu$m IRAS
surveys, along with some other information.  The first thing to notice
in this diagram is that we have reached flux limits almost 100 times
fainter than the deepest IRAS number counts at these wavelengths. We
are thus able to see much deeper into the universe than the IRAS
surveys and can provide considerably more powerful sampling of the
12$\mu$m galaxy population.

Secondly, our survey is the first flux limited 12$\mu$m
survey to be dominated by galaxies rather than stars. The deepest 
IRAS sample (Hacking \& Houck 1987) included
$\sim$50 objects of which only 5 were galaxies. As discussed above,
the present survey contains 50 objects above the 5$\sigma$ flux limit,
of which only 13 are stars. 

\begin{figure}
\resizebox{\hsize}{!}{\includegraphics{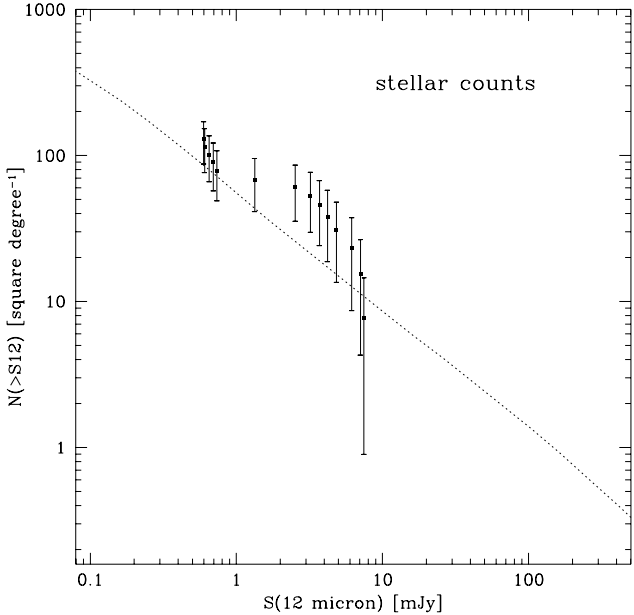}}
\caption{12$\mu$m Integral Counts for Stars.}
Same as in Fig.~\ref{intcounts}, but for the 13 sources identified as
stars in the 12$\mu$m survey. The fluxes are corrected by a factor
1/1.25 to account for the wide LW10 response function and for the
typical SEDs of stars.  A comparison is made
with the model stellar counts from Franceschini et al. (1991).
\label{starcounts}
\end{figure}

The integral counts of stellar identifications in the 12 $\mu$m survey
are shown in Fig.~\ref{starcounts}. We find good agreement with model
counts by Franceschini et al. (1991) based on the Bahcall and Soneira
galactic model and on a stellar luminosity function scaled from the V
band to $\lambda=12\ \mu m$ according to Hacking \& Houck (1987).
This agreement suggests that no major new stellar component is
emerging at faint fluxes with respect to those detected in the optical.

\begin{figure*}
\resizebox{\hsize}{!}{\includegraphics{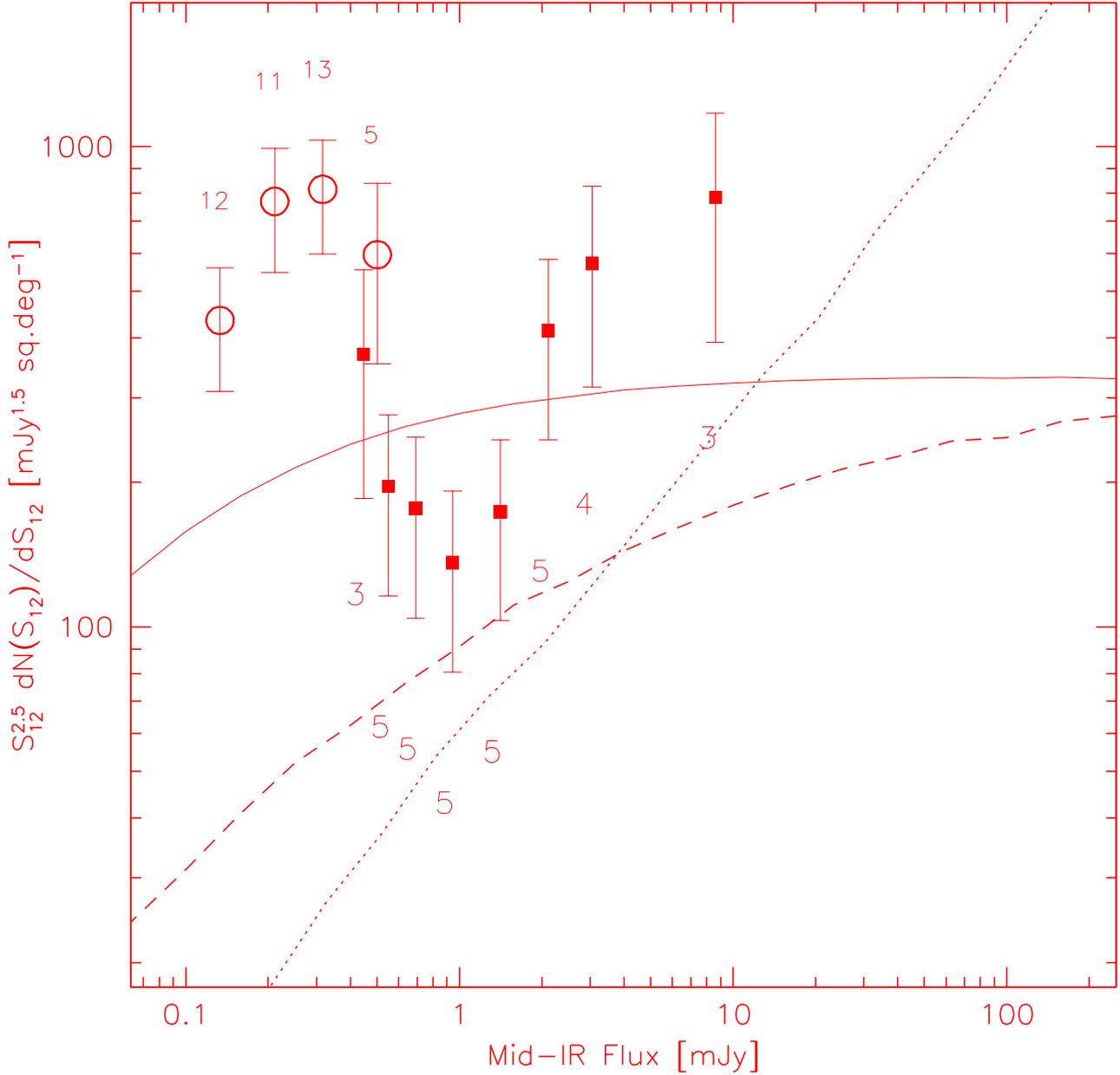}}
\caption{Euclidean-normalised Differential Number Counts for Galaxies
from the 12$\mu$m Survey}
The differential counts of sources identified with galaxies in the LW10 survey
are normalised here to the Euclidean law $S^{-2.5}$.
The filled squares are from our 12 $\mu$m survey, the open circles from
the 15 $\mu$m survey in the Hubble Deep Field (all with Poissonian error bars).
The number associated with each bin indicates the number of sources 
in the bin. The dashed line is the no-evolution curve, while the
continuous line 
corresponds to the model with evolution. The dotted line provides a comparison
with the expected stellar counts (see the corresponding integral counts in
Fig. ~\ref{intcounts}).
\label{difcounts}
\end{figure*}

We have so far shown the integral number counts for the galaxies in our survey.
A more statistically meaningful way to compare observed and theoretical 
number counts is to examine them in a
differential form, i.e. $dN(S_{12})/dS_{12}$
versus $S_{12}$. This is done in Fig.~\ref{difcounts}, where we report 
the Euclidean-normalised differential counts from our 12$\mu$m survey 
compared to the 15$\mu$m number counts derived from ISOCAM observations
of the Hubble Deep Field (Paper I). 
The lines correspond to predictions for Euclidean-normalised
counts based on both non-evolving and strongly evolving population models.

The no-evolution model is based on the local luminosity function
(Saunders et al., 1990) at 60$\mu$m and on the Rush et al (1993)
results at 12$\mu$m.  (for more details see Franceschini et
al. 1997). This minimal curve significantly under predicts the observed
counts from ISO. We thus appear to have detected evolution in the
12$\mu$m source population at a $3.5\sigma$ significance level.

On the other hand, our observed 12$\mu$m counts are matched by a model
assuming an evolving luminosity function. This is described in terms of
two populations, which we assume dominate the extragalactic sky
at these wavelengths:

(1) Gas-rich systems, i.e. spiral, irregular and starbursting galaxies, with
luminosity functions evolving with cosmic time as 
$N(L,z)=N(L,z=0)\ exp(k\ \tau (z))$, where $N(L,z=0)$ is the locally observed 
distribution (see above),
 $\tau (z)$ is the lookback time $(t_0 - t)/t_0$, where t is the age of
the universe at a redshift z and $t_0$ is the present age, and $k=3$ is the 
evolution parameter. This corresponds to density evolution, yielding an average
increase in galaxy co-moving number density of a factor of 5.8 at z=1 (for
an assumed $q_0$=0.15 value of the cosmological deceleration parameter).

(2) Active Galactic Nuclei, which are described by a model based again
on the Rush et al. (1993) local luminosity function and assuming pure
luminosity evolution: $N(L,z) = N(L_0,z)$, where $L(z) = L_0\ \exp(k
\tau [z])$ with $k=3$.

Note that the same model with the additional contribution of a
population of high-redshift starbursts (forming elliptical and S0
galaxies as described by Franceschini et al. 1994) accounts nicely for
the cosmological far-IR background recently detected in the
far-infrared and submillimeter wavebands by Puget et al. (1996),
Hauser et al. (1997b) and Fixsen et al (1998).

It is also interesting to compare the 12$\mu$m counts described here
with the 15$\mu$m counts from the HDF. These counts are derived from
two different ISOCAM filters (LW10 and LW3) with rather different
response functions. As can be seen in Fig. 6, there appears to be a
clear offset between the two differential galaxy counts by roughly a
factor 2 -- 4 (though the two bins with overlapping fluxes are in
formal agreement).  No simple model can explain this shift in the
counts by such a large factor over such a narrow flux interval. We
interpret this shift as probably not due to actual changes in the
counts, but to the different responses of the LW10 and LW3 filters to
the complex SEDs of galaxies in the mid-IR.
Specifically the 7$\mu$m PAH emission feature, which enters the LW3
15$\mu$m band at z$\sim$0.5 to 1, and the 10$\mu$m absorption
feature. Unfortunately, the strength of these mid-IR spectral features
varies considerably from object to object (see eg. Elbaz et al.,
1998).  A full understanding of the effects of these features on
mid-IR number counts thus awaits a better theoretical treatment, a
better understanding of the variation of these features locally, and a
better idea of the nature of objects making up the milli--Jansky
12$\mu$m source population. Xu et al. (1998) used a three component
model including cirrus, starburst and AGN contributions to fit the
mid-IR SEDs of a large sample of local
galaxies. They then extrapolate from this to predict the effects of
the mid-IR SEDs on number counts under various evolutionary
assumptions. Such an approach may be useful for understanding the
present work and its relation to the ISOHDF data. However, assumptions
would have to be made about the nature of the faint mid-IR galaxy
population, and whether it was significantly different from the local
galaxies studied in detail by IRAS and ISO. There are already
suggestions from the ISOHDF that there are more and bigger starbursts
in the faint mid-IR selected galaxies than in the local population
(Rowan-Robinson et al., 1997). At this stage we lack redshifts, and
thus luminosity and star-formation-rate estimates, for our 12$\mu$m
galaxies. A large number of assumptions would thus need to be made
about these objects for an empirical approach similar to Xu et
al. (1998) to be applied. There would thus be considerable
uncertainties in such an analysis.

A proper test of models of this population is thus even more
critically dependent on obtaining the redshifts of individual sources
than similar work at optical or far-IR wavelengths. We have therefore
begun a followup programme to identify and determine the
redshifts for all the 12$\mu$m galaxies discussed in this paper. Once
this data is available, we will be able to draw firmer conclusions
about the nature and evolution of the mJy 12$\mu$ source population.

Our number counts can directly set a lower limit to the extragalactic 
infrared background light between 8 and 15 $\mu$m. By integrating the 
light from the galaxies with a flux larger than 0.5 mJy, we find that
$\nu I_{\nu} ({\rm EBL _{12\mu m}}) > 0.50\pm 0.15 \, {\rm 
nWm^{-2}sr^{-1}}$. An upper limit of $468 {\rm nWm^{-2}sr^{-1}}$ has been 
reported by Hauser et al. (1998) from DIRBE (COBE) measurements, which 
are hampered by the zodiacal light.

\section{Conclusions}

We have performed a deep survey at 12$\mu$m using the CAM
instrument on the ISO satellite. We have detected 50 objects to a 5$\sigma$
flux threshold of $\sim 500\mu$Jy. in a 0.1 deg$^2$ area, of which 13
appear to be stars on the basis of optical images and optical-IR colours.
The remaining
37 objects appear to be galaxies. We have examined the source count
statistics for this population and find evidence for evolution in this
population, while for stars the counts are consistent with current
galactic structure models and extrapolations
of the optical luminosity functions to the mid-IR. 

Our galaxy counts, when compared with the deep ISOCAM counts 
in the Hubble Deep Field using the LW3 15$\mu$m filter,
also show evidence for significant effects from the complex
mid-infrared features
in the spectral energy distributions of galaxies.

\begin{acknowledgements}
It is a pleasure to thank Herve Aussel, David Elbaz, Matt Malkan and
Jean-Loup Puget for helpful comments and contributions. The Digitised
Sky Survey was produced at the STSCI under US Government Grant NAG
W--2166. This research has made use of the NASA/IPAC Extragalactic
Database (NED) which is operated by the Jet Propulsion Laboratory,
California Institute of Technology, under contract with the National
Aeronautics and Space Administration. The USNO-A survey was of considerable
help, and we would like to express our thanks to all those who helped to
produce it. DLC and ACB are
supported by an ESO Fellowship and by the EC TMR Network programme,
FMRX-CT96-0068.
\end{acknowledgements}

\end{document}